\documentclass{article}
\usepackage[utf8]{inputenc}
\usepackage{amsmath}
\usepackage{amssymb}
\usepackage{amsfonts}
\usepackage{esvect}
\usepackage{cite}
\usepackage{physics}
\usepackage{mathrsfs}
\usepackage{authblk}
\usepackage{geometry}
\usepackage{abstract}
\newtheorem{mypro}{Proposition}[section]

\title{\bf  Superintegrable systems from block separation of variables and unified derivation of their quadratic algebras}
\author{\large Zhe Chen\footnote{zhe.chen3@uqconnect.edu.au}, 
Ian Marquette\footnote{i.marquette@uq.edu.au}, 
and  Yao-Zhong Zhang\footnote{yzz@maths.uq.edu.au}}
\affil{School of Mathematics and Physics, The University of Queensland \\ Brisbane, QLD 4072, Australia}

\begin{document}

\maketitle
\begin{abstract}
We present a new method for constructing $D$-dimensional minimally superintegrable systems based on block coordinate separation of variables. We give two new families of superintegrable systems with $N$ ($N\leq D$) singular terms of the partitioned coordinates and involving arbitrary functions. These Hamiltonians generalize the singular oscillator and Coulomb systems. We derive their exact energy spectra via separation of variables. We also obtain the quadratic algebras satisfied by the integrals of motion of these models. We show how the quadratic symmetry algebras can be constructed by novel application of the gauge transformations from those of the non-partitioned cases. We demonstrate that these quadratic algebraic structures display an universal nature to the extent that their forms are independent of the functions in the singular potentials.
\end{abstract}

\section{Introduction}
A mechanical system is called ``superintegrable" \cite{mil} if it possesses more independent integrals of motion than its degrees of freedom. Superintegrable systems are interesting research topics in both classical and quantum theories due to their rich mathematical and physical properties. It has been shown that superintegrable models have connections to orthogonal polynomials and special functions, such as the Askey Scheme of orthogonal polynomials, exceptional orthogonal polynomials, hypergeometric functions, elliptic functions, Painlev\'e transcendents and higher-order analogs \cite{kal,mil,mil2,1,2,3,4}. Widely known examples of superintegrable models include the harmonic oscillator and hydrogen atom in three dimensions (as well as in $D$ dimensions). Many generalizations involving spins, magnetic fields and monopoles have also been discovered \cite{5,6,7,8,9,10,11,12}. An important property shared by superintegrable Hamiltonians is the existence of non-abelian symmetry algebras generated by their integrals of motion. In general, these symmetry algebras take the form of direct sums of higher rank Lie algebras and finitely generated polynomial algebras (which are non-linear generalizations of Lie algebras). Their structure constants can depend on central elements that can be Casimir operators of the higher rank Lie algebras. 

Two-dimensional superintegrable systems and their underlying quadratic algebras were classified in \cite{kal2,kal3,kal,mil,mil2}. Classification of three dimensional superintegrable Hamiltonians is a much more complicated problem \cite{13,14,15,16,17,das,cap}. To our knowledge, only for the so-called non-degenerate models a complete list of models is known and classification of semi-degenerate and degenerate is still ongoing. The related quadratic algebras remain to be systematically studied. 

Few families of $D$-dimensional superintegrable systems have been known in literature \cite{mil}. Such higher dimensional systems are expected to provide valuable insight into properties of superintegrable systems and their algebraic structures. Limited approaches have been developed to generate $D$-dimensional systems. Among them are the co-algebra and tensor product approaches, which allowed one to obtain models on curved spaces \cite{bal,gab}. Other approaches based on factorizations or intertwining relations \cite{sh10,ktv01} have also been used to generate $D$-dimensional superintegrable Hamiltonians \cite{lrbeh16,cnd09}. In the context of classical mechanics, families of subgroup separable superintegrable systems were introduced and studied in \cite{Kal10}. However, the derivation of symmetry algebras of $D$-dimensional superintegrable systems is in general a very difficult task even with the knowledge of other underlying algebraic structures such as the factorization or intertwining relations. There are high demands for new approaches. 

In \cite{LMZ} a direct approach was used to obtain the higher rank quadratic algebra for the $D$-dimensional model introduced in \cite{22}. A chain structure of quadratic algebras was discovered  in \cite{LMZ} and applied to derive the spectrum of the model. It was later demonstrated that such chain structures are in fact universal in the sense that they correspond to the algebras of integrals obtained by applying the co-algebra approach to the partial Casimir operators of $sl(2,\mathbb R)$ \cite{lat}. 

In this paper we introduce a new method based on block coordinate separation of variables to construct $D$-dimensional superintegrable systems. The connection between separation of variables and second order integrals of motion has played a role in studying Laplace and Helmholtz equations \cite{m84} as well as Schr\"odinger equation and Hamilton-Jacobi in curved spaces \cite{k86}. It was recognized as a way to classify superintegrable systems in 2D and 3D Euclidean spaces \cite{fmsuw65,svw67,e90}. This connection was also discussed in recent work \cite{kkm18}. The purpose of this work is to exploit blocks of coordinates and what will be referred as block separation of variables, where the $D$ coordinates in $D$ dimensions are partitioned into $N$ disjoint and non-empty blocks. Our models involve arbitrary functions of angles and the integrals of motion of the models display universal quadratic algebraic structures. To our knowledge, both our approach and models in this paper are new. We will present two classes of models. One is the so-called generalized $N$-singular harmonic oscillators. It is known that systems having double singular harmonic terms in the potential are also interesting due to their connection to monopole systems via Hurwitz transformations \cite{18,19,20}. In \cite{Faz}, a family of double singular oscillators was introduced and its quadratic algebra structure was obtained and used to derive the spectrum of the model. 
Another class of models we will introduce are referred to as generalized $N$-singular Coulomb systems that generalize models in \cite{21,22}.  We will present new families of superintegrable systems that involve arbitrary functions and include models in \cite{21,22} as special cases. All the models presented in this paper possess second and first order integrals of motion and are minimally superintegrable. Such systems have been much less studied in the literature than maximally superintegrable ones.

This work is organized as follows. In section 2, we introduce the generalized $N$ singular harmonic oscillators and obtain their energy spectrum and quadratic algebraic structure. In section 3, we introduce the generalized $N$ singular Coulomb systems and present their energy spectrum and derive their quadratic algebra structure,  generalizing the results of our recent work \cite{21}. At the end of this section we correct an error appeared in one of the quadratic algebra relations of \cite{21}.  We show how the symmetry algebras can be obtained from the models without partition of coordinates by identifying appropriate partial Casimir operators and gauge transformations. In section 4 we draw the conclusions of this work.

\section{Generalized $N$ singular harmonic oscillator}

\subsection{The model hamiltonian and energy spectrum}
Let $\{x_1,x_2,\cdots,x_{D} \}$ be the coordinates of the $D$-dimensional Euclidean space. We divide $D$ coordinates $\{x_1,x_2,\cdots,x_{D} \}$ into $N$ ($1 \leq N\leq D$) disjoint and nonempty blocks, say
\begin{equation}\label{par}
    \begin{aligned}
    &\mathcal{B}_1=\{x_{n_0+1}=x_1,\cdots,x_{n_1} \},~~~n_0=0,\\
    &\mathcal{B}_2=\{x_{n_1+1},\cdots,x_{n_2} \},\\
    &\cdots\\
    &\mathcal{B}_{i+1}=\{x_{n_i+1},\cdots,x_{n_{i+1}} \},\\
    &\cdots\\
    &\mathcal{B}_{N-1}=\{x_{n_{N-2}+1},\cdots,x_{n_{N-1}} \},\\
    &\mathcal{B}_{N}=\{x_{n_{N-1}+1},\cdots,x_{n_N}=x_{D} \},~~~n_N=D
    \end{aligned}
\end{equation}
and hereafter let
\begin{equation}
    d_i=|\mathcal{B}_i|=n_i-n_{i-1} \geq 1.
\end{equation}
Assuming the partition is non-trivial, we always have $d_i \geq 1$ and $N \geq 2$.

Now we propose the following family of quantum systems
\begin{equation}\label{sys1}
    \hat{H}=-\sum_{i=1}^{D}\frac{\partial^2}{\partial x_i^2}+\omega^2r^2+\frac{f_1(\Omega_1)}{x_1^2+\cdots+x_{n_1}^2}
+\frac{f_2(\Omega_2)}{x_{n_1+1}^2+\cdots+x_{n_2}^2}+\cdots+\frac{f_N(\Omega_N)}{x_{n_{N-1}+1}^2+\cdots+x_{n_{N}}^2},
\end{equation}
where $f_i(\Omega_i)$ are functions of $\Omega_i=\{\phi^{i}_{1},\cdots,\phi^{i}_{d_i-1} \}$ which is the set of angles from the spherical coordinates of individual block $\mathcal{B}_i$, i.e.,
\begin{equation}\label{blocksphere}
\begin{aligned}
    &x_{n_i}=r_i \cos \phi^{i}_{d_i-1}\\
    &x_{n_i-1}=r_i \sin \phi^{i}_{d_i-1} \cos \phi^{i}_{d_i-2}\\
    & \cdots \cdots \\
    &x_{n_{i-1}+1}=r_i \sin \phi^{i}_{d_i-1} \cdots \sin \phi^{i}_2\sin \phi^{i}_1.\\
\end{aligned}
\end{equation}
Introducing the notation
\begin{equation}\label{ri}
    r_i=\sqrt{\sum_{x\in \mathcal{B}_i}x^2},~~~i=1,2,\cdots,N,
\end{equation}
then (\ref{sys1}) can be expressed in the more compact form
\begin{equation}\label{sys}
    \hat{H}=-\sum_{i=1}^{D}\frac{\partial^2}{\partial x_i^2}+\omega^2r^2+\sum_{i=1}^{N}\frac{1}{r_i^2}f_i(\Omega_i).
\end{equation}
Note that in terms of $r_i$, we have $r=\sqrt{\sum_{i=1}^{D}x_i^2}=\sqrt{\sum_{i=1}^{N}r_i^2}$.

The eigenvalue problem $\hat{H}\Psi=E\Psi$ under coordinate system \eqref{blocksphere} can be written as
\begin{equation}
    \Psi=\prod_{i=1}^{N}r_i^{-(d_i-1)/2}R(r_1,\cdots,r_N)\prod_{i=1}^{N}Y_i(\Omega_i),
\end{equation}
where
\begin{equation}\label{fi}
    \left[-\hat{L}^2_i+f_i(\Omega_i)\right]Y_i(\Omega_i)=\lambda_i Y_i(\Omega_i)
\end{equation}
with 
\begin{equation}\label{L_i}
\hat{L}^2_i=\frac{1}{2}\sum_{x,y\in \mathcal{B}_i}\bigg(x\frac{\partial}{\partial y}-y\frac{\partial}{\partial x} \bigg)^2,
\end{equation}
and
\begin{equation}\label{radial}
    \bigg[-\sum_{i=1}^{N}\frac{\partial^2}{\partial r_i^2} +\omega^2r^2+\sum_{i=1}^{N}\frac{1}{r_i^2}(\lambda_i+\frac{1}{4}(d_i-1)(d_i-3))\bigg]R(r_1,\cdots,r_N)=ER(r_1,\cdots,r_N).
\end{equation}
In terms of $\lambda_i$'s, one can quote the result of Calogero three-body problem \cite{Cal} to get the formula for spectrum $E$
\begin{equation}\label{spectrum}
\begin{aligned}
    &E=2\omega \sum_{i=1}^{N}k_i+\omega \sum_{i=1}^{N}\gamma_i+\frac{\omega N}{2},~~~~~k_i=0,1,2,\cdots\\
    &\gamma_i=\frac{1}{2}(1+\sqrt{1+4\lambda_i+(d_i-1)(d_i-3)})~~~~~i=1,2,\cdots,N.
\end{aligned}
\end{equation}

{}From \eqref{fi}, it can be seen that the $\lambda_i$ is determined by $f_i(\Omega_i)$. Assuming each $f_i(\Omega_i)$ is of the form
\begin{equation}\label{fi1}
\begin{aligned}
    &f_i(\Omega_i)=F^i_{d_i-1}(\phi^{i}_{d_i-1},\cdots,\phi^{i}_1)\\
    &F^i_{d_i-1}(\phi^{i}_{d_i-1},\cdots,\phi^{i}_1)=G^{i}_{d_i-1}(\phi^{i}_{d_i-1})+\frac{1}{\sin^2 \phi^{i}_{d_i-1}}F^{i}_{d_i-2}(\phi^{i}_{d_i-2},\cdots,\phi^{i}_1)\\
    &F^{i}_{d_i-2}(\phi^{i}_{d_i-2},\cdots,\phi^{i}_1)=G^{i}_{d_i-2}(\phi^{i}_{d_i-2})+\frac{1}{\sin^2 \phi^{i}_{d_i-2}}F^{i}_{d_i-3}(\phi^{i}_{d_i-3},\cdots,\phi^{i}_1)\\
    & \cdots \cdots\\
    &F^i_2(\phi^i_2,\phi^i_1)=G^i_{2}(\phi^i_2)+\frac{1}{\sin^2 \phi^i_{2}}F^i_1(\phi^i_1)\\
    &F^i_1(\phi^i_1)=G^i_1(\phi^i_1),~~~i=1,2,\cdots,N,
\end{aligned}
\end{equation}
where $F^i_j$'s and $G^i_j$'s are functions. Then equation \eqref{fi} is completely separable, i.e. 
\begin{equation}
    Y_i(\Omega_i)=\prod_{k=1}^{d_i-1}h^i_k(\phi^i_k),~~~i=1,2,\cdots,N.
\end{equation}
where $h^i_j(\phi^i_j)$ satisfy the differential equations
\begin{equation}\label{hi}
\begin{aligned}
    &\bigg[-\frac{\partial^2}{\partial {\phi^i_j}^2}-(j-1)\cot \phi^i_j\frac{\partial }{\partial \phi^i_j}+G^i_j(\phi^i_j)+\frac{\alpha^i_{j-1}}{\sin^2 \phi^i_j}\bigg]h^i_j(\phi^i_j)=\alpha^i_{j}h^i_j(\phi^i_j),\\
     &i=1,2,\cdots,N~~~j=1,2,\cdots,d_i-1\\
    &\lambda_i=\alpha^i_{d_i-1},~~~\alpha^i_0=0.\\
\end{aligned}
\end{equation}

\vskip.1in
We now present two interesting special cases.
\vskip.2in
\noindent \underline{\bf\large Model 1:}
\vskip.1in
Let $f_i(\Omega_i)=\beta_i=$ real constants for $i=1,2,\cdots,N$, then Hamiltonian (\ref{sys}) reduces to the $N$-singular harmonic oscillator
\begin{equation}
    \hat{H}=-\sum_{i=1}^{D}\frac{\partial^2}{\partial x_i^2}+\omega^2 r^2+\frac{\beta_1}{x_1^2+\cdots+x^2_{n_1}}+\cdots+\frac{\beta_N}{x^2_{n_{N-1}+1}+\cdots+x_D^2}.
\end{equation}
This system is a generalization of the double singular harmonic oscillator considered in \cite{Faz}. For this model, the solution to \eqref{fi} is the spherical harmonics in $d_i$-dimensional space, with the corresponding eigenvalue $\lambda_i$ given by
\begin{equation}
    \lambda_i=l_i(l_i+d_i-2)+\beta_i,~~~l_i=0,1,2,\cdots,i=1,2,\cdots,N.
\end{equation}
The radial part $R(r_1,\cdots,r_N)$ is given by
\begin{equation}
\begin{aligned}
    &R(r_1,\cdots,r_N)=\prod_{i=1}^{N}r_i^{\gamma_i}\exp\big\{-\frac{\omega}{2}r_i^2 \big\}L^{(\gamma_i-1/2)}_{k_i}(\omega r_i^2),\\
    &\gamma_i=\frac{1}{2}(1+\sqrt{1+4l_i(l_i+d_i-2)+4\beta_i+(d_i-1)(d_i-3)}),
\end{aligned}
\end{equation}
where $L^{(\gamma_i-1/2)}_{k_i}$ is the Laguerre polynomial.

\vskip.2in
\noindent \underline{\bf\large Model 2:}
\vskip.1in
We set 
\begin{equation}\label{F11}
\begin{aligned}
    &F^1_1(\phi^1_1)=\frac{A(A-3)+B^2}{\cos^2 3\phi^1_1}-\frac{B(2A-3)\sin 3\phi^1_1}{\cos^2 3\phi^1_1}+9\bigg[\frac{2(2A-3)}{2A-3-2B\sin 3\phi^1_1}-\frac{2[(2A-3)^2-4B^2]}{(2A-3-2B\sin 3\phi^1_1)^2} \bigg]\\
    &G^1_2(\phi^1_2)=G^1_3(\phi^1_3)=\cdots=G^1_{d_i-1}(\phi^1_{d_i-1})=0
\end{aligned}
\end{equation}
and $f_2(\Omega_2)=f_3(\Omega_3)=\cdots=f_N(\Omega_N)=0$. Then the system becomes
\begin{equation}\label{osciexp}
    \hat{H}=-\sum_{i=1}^{D}\frac{\partial^2}{\partial x_i^2}+\omega^2 r^2+\frac{1}{x_1^2+x_2^2}F^1_1(\phi^1_1)
\end{equation}
Through some calculations, we obtain \cite{KYR}
\begin{equation}
    h^1_1(\phi^1_1)=(1-\sin 3\phi^1_1)^{(A-B)/6}(1+\sin 3\phi^1_1)^{(A+B)/6}\cdot \frac{1}{2A-3-2B\sin 3\phi^1_1}\hat{P}^{(\alpha,\beta)}_{l+1}(\sin 3\phi^1_1)
\end{equation}
where $\hat{P}^{(\alpha,\beta)}_{l+1}$ is the $X_1$ exceptional Jacobi polynomial \cite{GKM} and
\begin{equation}
    \alpha=\frac{A}{3}-\frac{B}{3}-\frac{1}{2},~~~\beta=\frac{A}{3}+\frac{B}{3}-\frac{1}{2}.
\end{equation}
The corresponding eigenvalue $\alpha^1_1$ is given by
\begin{equation}
    \alpha^1_1=(A+3J_1)^2,~~~J_1=0,1,2\cdots.
\end{equation}
The complete eigenfunction and eigenvalue to operator $-\hat{L}^2_1+f_1(\Omega)$ in \eqref{fi} is 
\begin{equation}
\begin{aligned}
    &Y_1(\Omega_1)=h^1_1(\phi^1_1)\cdot \prod_{a=2}^{d_1-1}(\sin \phi^1_a)^{c_{a-1}+\frac{1}{2}-\frac{a-1}{2}}P^{(c_{a-1},-1/2)}_{J_a}(\cos 2\phi^1_a)\\
    &c_a=\sum_{s=1}^{a}J_s+\frac{a-1}{2}+A+J_1,
\end{aligned}
\end{equation}
and
\begin{equation}
    \alpha^1_{d_1-1}=\bigg[2\sum_{a=1}^{d_1-1}J_a+\frac{d_1-2}{2}+A+J_1\bigg]^2-\frac{(d_1-2)^2}{4},~~~J_a=0,1,2,\cdots,
\end{equation}
where $P^{(c_{a-1},-1/2)}_{J_a}$ is the Jacobi polynomial.

\subsection{Integrals of motion and quadratic algebra structure}
Assuming each $f_i(\Omega_i)$ has the structure given by \eqref{fi1}. We find the following integrals of motion associated with each block $\mathcal{B}_i$ 
\begin{equation}\label{int1}
    \begin{aligned}
    &\hat{H}_i=-\sum_{x \in \mathcal{B}_i}\frac{\partial^2}{\partial x^2}+\omega^2 r_i^2+\frac{1}{r_i^2}f_i(\Omega_i)=-\nabla^2_i+\omega^2 r_i^2+\frac{1}{r_i^2}f_i(\Omega_i),\\
    &\hat{T}_i=\hat{L}^2_i-f_i(\Omega_i),\\
    &\hat{G}^i_j=\sum_{n_{i-1}+1\leq k<l}^{j}L_{kl}^2-\bigg(\sum_{k=n_{i-1}+1}^{j}x_k^2\bigg)\cdot \frac{1}{r_i^2}f_i(\Omega_i),\\
    &i=1,2,\cdots,N,~~~j=n_{i-1}+2,\cdots,n_i,
    \end{aligned}
\end{equation}
where
\begin{equation}
    L_{kl}=x_kp_l-x_lp_k,~~~p_a=\frac{\partial}{\partial x_a}.
\end{equation}
It can be seen that $\sum_{i=1}^{N}\hat{H}_i=\hat{H}$. The second set of integrals of motion is given by
\begin{equation}\label{int2}
    \hat{Z}_l=\sum_{1\leq i<j}^{n_l}L_{ij}^2-\bigg(\sum_{i=1}^{l}r_i^2 \bigg)\bigg(\sum_{i=1}^{l}\frac{1}{r_i^2}f_i(\Omega_i) \bigg),~~~l=2,\cdots,N.
\end{equation}
It should be understood that $\hat{G}^i_{d_i-1}=\hat{Z}_1=\hat{T}_1$. The integrals \eqref{int1} and \eqref{int2} define $D+N-1$ integrals of motion. When $N=2$ and $f_i(\Omega_i)$ are real constants, they reduce to the $D+1$ integrals obtained in \cite{Faz}. The above integrals satisfy
\begin{equation}
    \comm{\hat{G}^i_j}{\hat{T}_k}=\comm{\hat{T}_p}{\hat{H}_q}=\comm{\hat{H}_m}{\hat{G}^{n}_l}=\comm{\hat{G}^a_b}{\hat{G}^c_d}=0,
\end{equation}
\begin{equation}
    \comm{\hat{Z}_i}{\hat{Z}_j}=0=\comm{\hat{Z}_m}{\hat{H}_n},~~~m<n
\end{equation}
and
\begin{equation}
    \comm{\hat{Z}_l}{\sum_{i=1}^{l}\hat{H}_i}=0=\comm{\hat{Z}_k}{\hat{G}^i_j}
\end{equation}
Moreover we can show that for given $l$, $\hat{Z}_l,\hat{H}_l$ satisfy the following quadratic algebra relations:
\begin{equation}\label{quadratic-alg1} 
    \begin{aligned}
    \comm{\hat{Z}_l}{\hat{H}_l}=&\hat{Y}_l,\\
    \comm{\hat{Z}_l}{\hat{Y}_l}=&+8\big[\hat{Z}_l-\frac{1}{4}(D_l-2)^2 \big]\big(\sum_{i=1}^{l}\hat{H}_i\big)-8\acomm{\hat{Z}_l-\frac{1}{4}(D_l-2)^2}{\hat{H}_l}\\
    &+8\bigg[-\hat{Z}_{l-1}+\hat{T}_l+\frac{1}{4}(D_{l-1}-2)^2-\frac{1}{4}(d_l-2)^2+1 \bigg]\big(\sum_{i=1}^{l}\hat{H}_i\big)-16\hat{H}_l,\\
    \comm{\hat{H}_l}{\hat{Y}_l}=&-8\big(\sum_{i=1}^{l}\hat{H}_i\big)\hat{H}_l+8\hat{H}_l^2-16\omega^2\big[\hat{Z}_l-\frac{1}{4}(D_l-2)^2 \big]\\
    &-8\omega^2\bigg[-2\hat{Z}_{l-1}-2\hat{T}_l+\frac{1}{2}(D_{l-1}-1)(D_{l-1}-3)+\frac{1}{2}(d_l-1)(d_l-3)-1 \bigg].
    \end{aligned}
\end{equation}
These quadratic algebra relations can be obtained by applying the proposition (\ref{pop}) in the Appendix. This is seen as follows. 

For fixed $l$, 
\begin{equation}\label{H}
\begin{aligned}
    e^{-W}\bigg(\sum_{i=1}^{l}\hat{H}_i\bigg)e^W=&-\frac{\partial^2}{\partial r'^2}-\frac{\partial^2}{\partial r_l^2}+\omega^2(r'^2+r_l^2)+\frac{1}{r'^2}\big[-\hat{Z}_{l-1}+\frac{1}{4}(D_{l-1}-1)(D_{l-1}-3) \big]\\
    &+\frac{1}{r_l^2}\big[-\hat{T}_l+\frac{1}{4}(d_l-1)(d_l-3) \big],
\end{aligned}
\end{equation}
where $W$, $D_{l-1}$ and $r'$ are given by
\begin{equation}
\begin{aligned}
    W=-\frac{D_{l-1}-1}{2}\log r'-\frac{d_l-1}{2}\log r_l,~~~D_{l-1}=\sum_{i=1}^{l-1}d_i,~~~r'=\sqrt{\sum_{i=1}^{l-1}r_i^2}.
\end{aligned}
\end{equation}
Similarly we also find
\begin{equation}\label{Z}
\begin{aligned}
    &e^{-W}\hat{Z}_l e^W-\frac{1}{4}(D_l-2)^2\\
    &~~~~=(r'^2+r_l^2)\bigg(\frac{\partial^2}{\partial r'^2}+\frac{\partial^2}{\partial r_l^2} \bigg)-\bigg(r'\frac{\partial}{\partial r'}+r_l\frac{\partial}{\partial r_l} \bigg)^2\\
    &~~~~~~-(r'^2+r_l^2)\bigg\{\frac{1}{r'^2}\big[-\hat{Z}_{l-1}+\frac{1}{4}(D_{l-1}-1)(D_{l-1}-3) \big]
    +\frac{1}{r_l^2}\big[-\hat{T}_l+\frac{1}{4}(d_l-1)(d_l-3) \big]\bigg\}\\
\end{aligned}
\end{equation}
\begin{equation}\label{H2}
    e^{-W}\hat{H}_le^W=-\frac{\partial^2}{\partial r_l^2}+\omega^2 r_l^2+\frac{1}{r_l^2}\big[-\hat{T}_l+\frac{1}{4}(d_l-1)(d_l-3) \big].
\end{equation}
Comparing \eqref{H}, \eqref{Z} and \eqref{H2} with the operators $\mathcal{H}$, $\mathcal{Z}$ and $\mathcal{H}_2$ given in proposition \eqref{pop}, we see that because $\hat{Z}_{l-1}$ and $\hat{T}_l$ mutually commute and moreover commute with any differential operators and functions of $r'$ and $r_l$, they may be identified with the constants $g_1$ and $g_2$, respectively. Therefore, the operators in \eqref{H}, \eqref{Z} and \eqref{H2} should satisfy the same relations as those given in proposition \eqref{pop} for  $\mathcal{H}$, $\mathcal{Z}$ and $\mathcal{H}_2$, with $g_1, g_2$ replaced by $\hat{Z}_{l-1}$ and $\hat{T}_l$, respectively. This leads to the quadratic algebra (\ref{quadratic-alg1}).

\section{Generalized $N$ singular Coulomb system}
In this section, we introduce a family of superinetgrable systems in the $D$-dimensional Euclidean space involving Coulomb and $N$ singular inverse square potentials. These systems can be regarded as extensions of those presented in \cite{21}.

\subsection{The model hamiltonian and energy spectrum}
Consider the same partition as that given in \eqref{par}. We propose the generalized $N$ singular Coulomb system hamiltonian, 
\begin{equation}\label{sys2a}
    \hat{H}_{coul}=-\sum_{i=1}^{D}\frac{\partial^2}{\partial x_i^2}-\frac{\eta}{r}+\frac{g_1(\Omega_1)}{x_1^2+\cdots+x_{n_1}^2}
+\frac{g_2(\Omega_2)}{x_{n_1+1}^2+\cdots+x_{n_2}^2}+\cdots+\frac{g_{N-1}(\Omega_{N-1})}{x_{n_{N-2}+1}^2+\cdots+x_{n_{N-1}}^2},
\end{equation}
where $g_i(\Omega_i)$ are functions of the sets of angles $\Omega_i=\{\phi^{i}_{1},\cdots,\phi^{i}_{d_i-1} \}$ in \eqref{blocksphere}. In terms of the notation introduced in the previous section, the above hamiltonian can be written as
\begin{equation}\label{sys2}
    \hat{H}_{coul}=-\sum_{i=1}^{D}\frac{\partial^2}{\partial x_i^2}-\frac{\eta}{r}+\sum_{i=1}^{N-1}\frac{1}{r_i^2}g_i(\Omega_i).
\end{equation}
This hamiltonian reduces to that proposed in \cite{21} when $g_i(\Omega_i)=\alpha_i=$ real constants.

By means of the spherical coordinates \eqref{blocksphere} in each block, the eigenvalue problem $\hat{H}_{coul} \Psi=E\Psi$ can be written as
\begin{equation}
    \Psi=\prod_{i=1}^{N}r_i^{-(d_i-1)/2}R(r_1,\cdots,r_N)\prod_{i=1}^{N}Y_i(\Omega_i),
\end{equation}
\begin{equation}\label{coulang}
\begin{aligned}
    &\bigg[-\hat{L}^2_i+g_i(\Omega_i)+\frac{1}{4}(d_i-1)(d_i-3)\bigg]Y_i(\Omega_i)=\lambda_i Y_i(\Omega_i),~~~i=1,2,\cdots,N-1,\\
    &\bigg[-\hat{L}^2_N+\frac{1}{4}(d_N-1)(d_N-3)\bigg]Y_N(\Omega_N)=\lambda_NY_N(\Omega_N),
\end{aligned}
\end{equation}
where $L_i^2$ is given by (\ref{L_i}), and
\begin{equation}
    \bigg[-\sum_{i=1}^{N}\frac{\partial^2}{\partial r_i^2}  -\frac{\eta}{r}+\sum_{i=1}^{N}\frac{\lambda_i}{r_i^2}\bigg]R(r_1,\cdots,r_N)=ER(r_1,\cdots,r_N).
\end{equation}

In the following, we will assume $g_i(\Omega_i)$ has the same structure as in \eqref{fi1}, i.e. $g_i(\Omega_i)=F^i_{d_i-1}(\phi^{i}_{d_i-1},\cdots,\phi^{i}_1)$ with $F^i_{d_i-1}$ given by \eqref{fi1}. Then equation \eqref{coulang} is completely separable, i.e.,
\begin{equation}
    Y_i(\Omega_i)=\prod_{k=1}^{d_i-1}h^i_k(\phi^i_k),~~~i=1,2,\cdots,N,
\end{equation}
where $h^i_j(\phi^i_j)$ satisfy differential equations similar to those in (\ref{hi}).
In terms of $\lambda_i$'s, using the result in \cite{21}, the eigenvalue of $\hat{H}_{coul}$ is given by
\begin{equation}\label{couleigen}
    E=-\frac{\eta^2}{(2N_r+4\sum_{s=1}^{N-1}J_s+2N-1+2\sum_{s=1}^{N}\gamma_s)^2},~~~N_r=0,1,\cdots.
\end{equation}
where
\begin{equation}
    \gamma_j=\frac{\sqrt{1+4\lambda_j}}{2},~~~j=1,2,\cdots,N.
\end{equation}
The radial wave function $R(r_1,\cdots,r_N)$ is 
\begin{equation}
    R(r_1,\cdots,r_{N})=r^{-\frac{N-1}{2}}F(r)\prod_{i=1}^{N-1}y_i(\theta_i),
\end{equation}
where $F(r)$ can be written in terms of the Laguerre polynomial $L^{(2\kappa-1)}_{N_r}$ as
\begin{equation}
    \begin{aligned}
    &F(r)=r^{\kappa}e^{-\sqrt{-E}r}L^{(2\kappa-1)}_{N_r}(2\sqrt{-E}r),~~~\kappa=2\sum_{s=1}^{N-1}J_s+N-\frac{1}{2}+\frac{1}{2}\sum_{s=1}^{N}\sqrt{1+4\lambda_s},\\
    \end{aligned}
\end{equation}
and
\begin{equation}\label{coulsol}
\begin{aligned}
    &y_i(\theta_i)=(\sin \theta_i)^{\kappa_{i-1}+1-\frac{i}{2}} (\cos \theta_i)^{\gamma_{i+1}+\frac{1}{2}}P^{(\kappa_{i-1},\gamma_{i+1})}_{J_i}(\cos 2\theta_i),\\
    &\kappa_{i}=2\sum_{s=1}^{i}J_s+i+\sum_{s=1}^{i+1}\gamma_s,~~~\gamma_j=\frac{\sqrt{1+4\lambda_j}}{2},~~~b_i=\kappa_i^2-\frac{(i-1)^2}{4},\\
    &i=1,2,\cdots,N-1,~~~j=1,2,\cdots,N,~~~J_s=0,1,\cdots,\\
\end{aligned}
\end{equation}
with $\theta_i$  being angles in the spherical coordinates of $r_1,\cdots,r_N$:
\begin{equation}
\begin{aligned}
    &r_N=r\cos \theta_{N-1}\\
    &r_{N-1}=r\sin \theta_{N-1} \cos \theta_{N-2}\\
    &\cdots \cdots\\
    &r_2=r\sin \theta_{N-1}\sin \theta_{N-2} \cdots \sin \theta_2 \cos \theta_1\\
    &r_1=r\sin \theta_{N-1}\sin \theta_{N-2} \cdots \sin \theta_2 \sin \theta_1.
\end{aligned}
\end{equation}

\vskip.2in
We now present two special cases.
\vskip.2in
\noindent \underline{\bf\large Model 1:}
\vskip.1in
Let $g_i(\Omega)=\alpha_i=$ constant, $i=1,2,\cdots,N-1$. We then recover the system proposed in \cite{21},
\begin{equation}
    \hat{H}_{coul}=-\sum_{i=1}^{D}\frac{\partial^2}{\partial x_i^2}-\frac{\eta}{r}+\frac{\alpha_1}{x_1^2+\cdots+x^2_{n_1}}+\cdots+\frac{\alpha_{N-1}}{x^2_{n_{N-2}+1}+\cdots+x^2_{n_{N-1}}},
\end{equation}
which in turn reduces to the model studied in \cite{22} when $N=D$.
\vskip.2in
\noindent \underline{\bf\large Model 2:}
\vskip.1in
Assuming $F^1_1$ has the same form as in (\ref{F11}) and $g_2(\Omega_2)=g_3(\Omega_3)=\cdots=g_{N-1}(\Omega_{N-1})=0$,
then we obtain 
\begin{equation}
    \hat{H}_{coul}=-\sum_{i=1}^{D}\frac{\partial^2}{\partial x_i^2}-\frac{\eta}{r}+\frac{1}{x_1^2+x_2^2}F^1_1(\phi^1_1),
\end{equation}
where $F^1_1(\phi^1_1)$ is given by (\ref{F11}).
The solution to the angular part is again given by the  $X_1$ exceptional Jacobi polynomial as follows
\begin{equation}
\begin{aligned}
    &Y_1(\Omega_1)=h^1_1(\phi^1_1)\cdot \prod_{a=2}^{d_1-1}(\sin \phi^1_a)^{c_{a-1}+\frac{1}{2}-\frac{a-1}{2}}P^{(c_{a-1},-1/2)}_{J_a}(\cos 2\phi^1_a)\\
    &c_a=\sum_{s=1}^{a}J_s+\frac{a-1}{2}+A+J_1,
\end{aligned}
\end{equation}
and
\begin{equation}
    \alpha^1_{d_1-1}=\bigg[2\sum_{a=1}^{d_1-1}J_a+\frac{d_1-2}{2}+A+J_1\bigg]^2-\frac{(d_1-2)^2}{4},~~~J_a=0,1,2,\cdots,
\end{equation}
where $P^{(c_{a-1},-1/2)}_{J_a}$ is the Jacobi polynomial and $h^1_1(\phi^1_1)$ is
\begin{equation}
    h^1_1(\phi^1_1)=(1-\sin 3\phi^1_1)^{(A-B)/6}(1+\sin 3\phi^1_1)^{(A+B)/6}\cdot \frac{1}{2A-3-2B\sin 3\phi^1_1}\hat{P}^{(\alpha,\beta)}_{l+1}(\sin 3\phi^1_1)
\end{equation}
\begin{equation}
    \alpha=\frac{A}{3}-\frac{B}{3}-\frac{1}{2},~~~\beta=\frac{A}{3}+\frac{B}{3}-\frac{1}{2}.
\end{equation}
The corresponding eigenvalue $\alpha^1_1$ is given by
\begin{equation}
    \alpha^1_1=(A+3J_1)^2,~~~J_1=0,1,2\cdots.
\end{equation}

\subsection{Integrals of motion and quadratic algebraic structure}
Assuming each $g_i(\Omega_i),~i=1,2,\cdots, N-1,$ has the structure given by \eqref{fi1}. The integrals of motion are given by 
\begin{equation}
    \hat{T}_i=\hat{L}^2_i-g_i(\Omega_i),~~~~~~\hat{T}_N=\hat{L}^2_N\\
\end{equation}
\begin{equation}
    \hat{Z}_l=\sum_{1\leq i<j}^{n_l}L_{ij}^2-\bigg(\sum_{a=1}^{l}r_a^2 \bigg)\bigg(\sum_{a=1}^{l}\frac{1}{r_a^2}g_a(\Omega_a) \bigg),~~~l=2,\cdots,N-1.
\end{equation}
\begin{equation}
    \hat{X}_i=\sum_{a=1}^{D}\acomm{L_{ia}}{p_a}+\frac{\eta}{r}x_i-2x_i\cdot \sum_{a=1}^{N-1}\frac{1}{r_a^2}g_{a}(\Omega_a),~~~i=n_{N-1}+1,\cdots,D.
\end{equation}
The operators $\hat{X}_i$'s are obtained by using the proposition $2.1$ in \cite{21}. 
We also have the following integrals of motion
\begin{equation}
    \begin{aligned}
    &\hat{S}_l=\sum_{1\leq i<j}^{l}L_{ij}^2-\bigg(\sum_{a=1}^{l}x_a^2 \bigg)\bigg(\sum_{a=1}^{N-1}\frac{1}{r_a^2}g_a(\Omega_a) \bigg),~~~l=n_{N-1}+1,\cdots D-1,\\
    &\hat{Y}_p=\sum_{n_{p-1}+1\leq i<j}^{D}L_{ij}^2-\bigg(\sum_{a=p}^{N}r_a^2 \bigg)\bigg(\sum_{a=p}^{N-1}\frac{1}{r_a^2}g_a(\Omega_a) \bigg),~~~p=1,2,\cdots,N-1\\
    &\hat{J}_p=\sum_{p \leq i<j}^{D}L_{ij}^2,~~~~p=n_{N-1}+1,\cdots,D-1.
    \end{aligned}
\end{equation}
Notice that 
\begin{equation}
   \hat{J}_{n_{N-1}+1}=\hat{T}_N=\hat{L}^2_N.
\end{equation}
These integrals satisfy
\begin{equation}
\begin{aligned}
    \comm{\hat{Z}_p}{\hat{Z}_l}=0=\comm{\hat{S}_i}{\hat{Z}_j},~~~\comm{\hat{Y}_i}{\hat{Y}_j}=0=\comm{\hat{J}_k}{\hat{Y}_l},~~~\comm{\hat{Y}_1}{\hat{Z}_i}=0
\end{aligned}
\end{equation}
In what follows, we present the quadratic algebra relations among the integrals. 
We define the numbers
\begin{equation}
    \begin{aligned}
     &\mathcal{N}_p=\bigg(\sum_{i=1}^{p}d_i-2\bigg)\bigg(\sum_{i=1}^{p}\frac{d_i-1}{2} \bigg)-\bigg(\sum_{i=1}^{p}\frac{d_i-1}{2} \bigg)^2,~~p=2,\cdots,N-1,\\
    &\mathcal{M}_p=\bigg(\sum_{i=p}^{N}d_i-2\bigg)\bigg(\sum_{i=p}^{N-1}\frac{d_i-1}{2} \bigg)-\bigg(\sum_{i=p}^{N-1}\frac{d_i-1}{2} \bigg)^2,~~~p=1,\cdots,N-1,\\
    &\mathcal{U}_p=(p-2)\bigg(\sum_{i=1}^{N-1}\frac{d_i-1}{2} \bigg)-\bigg(\sum_{i=1}^{N-1}\frac{d_i-1}{2} \bigg)^2,~~~p=n_{N-1}+1,\cdots,D-1,\\
\end{aligned}
\end{equation}
$\hat{X}_D$ and $\hat{Z}_N=\hat{Y}_1$ satisfy following commutation relations
\begin{equation}\label{YX}
    \begin{aligned}
    &\comm{\hat{Y}_1}{\hat{X}_D}=\hat{W}_D,\\
    &\comm{\hat{Y}_1}{\hat{W}_D}=-2\acomm{\hat{Y}_1}{\hat{X}_D}+(D-1)(D-3)\hat{X}_D,\\
    &\comm{\hat{X}_D}{\hat{W}_D}=2\hat{X}_D^2-8(\hat{S}_{D-1}-\mathcal{U}_{D-1})\hat{H}_{coul}+16(\hat{Y}_1-\mathcal{M}_1)\hat{H}_{coul}-2(N+d_N-2)^2\hat{H}_{coul}-2\eta^2.
    \end{aligned}
\end{equation}
Notice that for permutation operator $\sigma_{jD}$ interchanging indices $j$, $D$, where $j\in\{D,D-1,\cdots,D-d_N+1 \}$, one can show
\begin{equation}
    \sigma_{jD}\circ \hat{X}_D \circ \sigma^{-1}_{jD}=\hat{X}_j,~~~\sigma_{jD}\circ \hat{Y}_1 \circ \sigma^{-1}_{jD}=\hat{Y}_1,~~~ \sigma_{jD}\circ \hat{H}_{coul} \circ \sigma^{-1}_{jD}=\hat{H}_{coul}.
\end{equation}
Applying $\sigma_{jD}$ on both sides of \eqref{YX}, we have
\begin{equation}
    \begin{aligned}
    &\comm{\hat{Y}_1}{\hat{X}_j}=\hat{W}_j\\
    &\comm{\hat{Y}_1}{\hat{W}_j}=-2\acomm{\hat{Y}_1}{\hat{X}_j}+(D-1)(D-3)\hat{X}_j\\
    &\comm{\hat{X}_j}{\hat{W}_j}=2\hat{X}_j^2-8(\sigma_{jD}\circ \hat{S}_{D-1}\circ \sigma^{-1}_{jD}-\mathcal{U}_{D-1})\hat{H}_{coul}+16(\hat{Y}_1-\mathcal{M}_1)\hat{H}_{coul}-2(N+d_N-2)^2\hat{H}_{coul}-2\eta^2,
    \end{aligned}
\end{equation}
where $\sigma_{jD}\circ \hat{S}_{D-1}\circ \sigma^{-1}_{jD}$ is another integral of motion and is explicitly given by
\begin{equation}\label{correction}
    \begin{aligned}
    \sigma_{jD}\circ \hat{S}_{D-1}\circ \sigma^{-1}_{jD}=&(r^2-x_j^2)\bigg(\sum_{i=1}^{D}\frac{\partial^2}{\partial x_i^2}-\frac{\partial^2}{\partial x_j^2} \bigg)-\bigg(\sum_{i=1}^{D}x_i\frac{\partial}{\partial x_i}-\frac{\partial}{\partial x_j} \bigg)^2\\
    &-(D-3)\bigg(\sum_{i=1}^{D}x_i\frac{\partial}{\partial x_i}-\frac{\partial}{\partial x_j} \bigg)-(r^2-x_j^2)\cdot \bigg(\sum_{a=1}^{N-1}\frac{1}{r_a^2}g_a(\Omega_a) \bigg).
    \end{aligned}
\end{equation}

By computations similar to those presented in \cite{21}, we can obtain the quadratic commutation relations for $\hat{Z}_p$ and $\hat{Y}_p$, $~2\leq p\leq N-1$,
\begin{equation}
\begin{aligned}
    \comm{\hat{Z}_p}{\comm{\hat{Z}_p}{\hat{Y}_p}}=&-8(\hat{Z}_p-\mathcal{N}_p)^2-8 \acomm{\hat{Z}_p-\mathcal{N}_p}{\hat{Y}_p-\mathcal{M}_p}\\
    &-4\bigg[(p-2)(N+d_N-1)-p^2+p+4+2\bigg(-\hat{T}_p+\frac{1}{4}(d_p-1)(d_p-3) \bigg) \bigg](\hat{Z}_p-\mathcal{N}_p)\\
    &+4(N+d_N-p)(N+d_N-p-4)(\hat{Y}_p-\mathcal{M}_p)+8(\hat{Y}_1-\mathcal{M}_1+\hat{Z}_{p-1}-\mathcal{N}_{p-1})(\hat{Z}_p-\mathcal{N}_p)\\
    &-4\bigg[N+d_N-p-4+2\bigg(-\hat{T}_p+\frac{1}{4}(d_p-1)(d_p-3) \bigg) \bigg](\hat{Y}_1-\mathcal{M}_1)\\
    &-4\bigg[(N+d_N-p-1)(N+d_N-p-4)-2\bigg(-\hat{T}_p+\frac{1}{4}(d_p-1)(d_p-3) \bigg) \bigg](\hat{Z}_{p-1}-\mathcal{N}_{p-1})\\
    &+4(N+d_N-p)(N+d_N-5)\bigg(-\hat{T}_p+\frac{1}{4}(d_p-1)(d_p-3) \bigg)\\
    &+4(p-1)(N+d_N-p)(\hat{Y}_{p+1}-\mathcal{M}_{p+1})-8(\hat{Y}_1-\mathcal{M}_1)(\hat{Y}_{p+1}-\mathcal{M}_{p+1})\\
    &+8(\hat{Z}_p-\mathcal{N}_p)(\hat{Y}_{p+1}-\mathcal{M}_{p+1})+8(\hat{Z}_{p-1}-\mathcal{N}_{p-1})(\hat{Y}_{p+1}-\mathcal{M}_{p+1}),
\end{aligned}
\end{equation}
\begin{equation}
\begin{aligned}
    \comm{\hat{Y}_p}{\comm{\hat{Z}_p}{\hat{Y}_p}}=&+8(\hat{Y}_p-\mathcal{M}_p)^2+8\acomm{\hat{Z}_p-\mathcal{N}_p}{\hat{Y}_p-\mathcal{M}_p}-4p(p-4)(\hat{Z}_p-\mathcal{N}_p)\\
    &+4\bigg[(p-2)(N+d_N-1)-p^2+p+4+2\bigg(-\hat{T}_p+\frac{1}{4}(d_p-1)(d_p-3) \bigg) \bigg](\hat{Y}_1-\mathcal{M}_p)\\
    &-8(\hat{Z}_{p-1}-\mathcal{N}_{p-1})(\hat{Y}_p-\mathcal{M}_p)-8(\hat{Y}_{1}-\mathcal{M}_1)(\hat{Y}_p-\mathcal{M}_p)\\
    &+4\bigg[p-4+2\bigg(-\hat{T}_p+\frac{1}{4}(d_p-1)(d_p-3) \bigg) \bigg](\hat{Y}_1-\mathcal{M}_1)+8(\hat{Z}_{p-1}-\mathcal{N}_{p-1})(\hat{Y}_1-\mathcal{M}_1)\\
    &-4p(N+d_N-p-1)(\hat{Z}_{p-1}-\mathcal{N}_{p-1})-4p(N+d_N-5)\bigg(-\hat{T}_p+\frac{1}{4}(d_p-1)(d_p-3) \bigg)\\
    &+4\bigg[(p-4)(p-1)-2\bigg(-\hat{T}_p+\frac{1}{4}(d_p-1)(d_p-3) \bigg) \bigg](\hat{Y}_{p+1}-\mathcal{M}_{p+1})\\
    &-8(\hat{Z}_{p-1}-\mathcal{N}_{p-1})(\hat{Y}_{p+1}-\mathcal{M}_{p+1})-8(\hat{Y}_{p+1}-\mathcal{M}_{p+1})(\hat{Y}_p-\mathcal{M}_p).
\end{aligned}
\end{equation}

For $\hat{S}_p$ and $\hat{J}_p$, we have
\begin{equation}
\begin{aligned}
    \comm{\hat{S}_p}{\comm{\hat{S}_p}{\hat{J}_p}}=&-8(\hat{S}_p-\mathcal{U}_p)^2-8 \acomm{\hat{S}_p-\mathcal{U}_p}{\hat{J}_p}\\
    &-4\bigg[(p+N+d_N-D-3)(N+d_N-1)-(p+N+d_N-D-1)^2+p+N+d_N-D+3 \bigg](\hat{S}_p-\mathcal{U}_p)\\
    &+4(D-p+1)(D-p-3)\hat{J}_p+8(\hat{Y}_1-\mathcal{M}_1+\hat{S}_{p-1}-\mathcal{U}_{p-1})(\hat{S}_p-\mathcal{U}_p)\\
    &-4(D-p-3)(\hat{Y}_1-\mathcal{M}_1)-4(D-p)(D-p-3)(\hat{S}_{p-1}-\mathcal{U}_{p-1})\\
    &+4(p+N+d_N-D-2)(D-p+1)\hat{J}_{p+1}-8(\hat{Y}_1-\mathcal{M}_1)\hat{J}_{p+1}\\
     &+8(\hat{S}_p-\mathcal{U}_p)\hat{Y}_{p+1}+8(\hat{Z}_{p-1}-\mathcal{N}_{p-1})\hat{J}_{p+1},
\end{aligned}
\end{equation}
\begin{equation}
\begin{aligned} 
    \comm{\hat{J}_p}{\comm{\hat{S}_p}{\hat{J}_p}}=&+8\hat{J}_p^2+8\acomm{\hat{S}_p-\mathcal{U}_p}{\hat{J}_p}-4(p+N+d_N-D-1)(p+N+d_N-D-5)(\hat{S}_p-\mathcal{U}_p)\\
    &+4\bigg[(p+N+d_N-D-3)(N+d_N-1)-(p+N+d_N-D-1)^2+p+N+d_N-D+3 \bigg](\hat{Y}_p-\mathcal{M}_p)\\
    &-8(\hat{S}_{p-1}-\mathcal{U}_{p-1})\hat{J}_p-8(\hat{Y}_1-\mathcal{M}_1)\hat{J}_p\\
    &+4(p+N+d_N-D-5)(\hat{Y}_1-\mathcal{M}_1)+8(\hat{S}_{p-1}-\mathcal{U}_{p-1})(\hat{Y}_1-\mathcal{M}_1)\\
    &-4(p+N+d_N-D-1)(D-p)(\hat{S}_{p-1}-\mathcal{U}_{p-1})\\
    &+4(p+N+d_N-D-5)(p+N+d_N-D-2)\hat{J}_{p+1}\\
    &-8(\hat{S}_{p-1}-\mathcal{U}_{p-1})\hat{J}_{p+1}-8\hat{J}_{p+1}\hat{J}_p.
\end{aligned}
\end{equation}

Two remarks are in order. 1. When $g_i(\Omega_i)=\alpha_i=$ constants, the above quadratic algebra relations reduce to those given in \cite{21}; 2. There is an error in the 3rd relation of eq.(54) of \cite{21}:  namely, $\hat{Z}_{N-1}$ on the right hand side of that relation should be replaced by $\sigma_{jD}\circ \hat{S}_{D-1}\circ \sigma^{-1}_{jD}$ with [c.f. (\ref{correction}) above]
\begin{equation}
    \begin{aligned}
    \sigma_{jD}\circ \hat{S}_{D-1}\circ \sigma^{-1}_{jD}=&(r^2-x_j^2)\bigg(\sum_{i=1}^{D}\frac{\partial^2}{\partial x_i^2}-\frac{\partial^2}{\partial x_j^2} \bigg)-\bigg(\sum_{i=1}^{D}x_i\frac{\partial}{\partial x_i}-\frac{\partial}{\partial x_j} \bigg)^2\\
    &-(D-3)\bigg(\sum_{i=1}^{D}x_i\frac{\partial}{\partial x_i}-\frac{\partial}{\partial x_j} \bigg)-(r^2-x_j^2)\cdot \bigg(\sum_{a=1}^{N-1}\frac{\alpha_a}{r_a^2} \bigg).
    \end{aligned}
\end{equation}

\section{Conclusion}
In this work, we have developed an approach based on block coordinate separation of variables and applied it to construct $D$-dimensional quantum superintegrable Hamiltonians and their higher rank quadratic algebras. These models involve arbitrary functions which correspond to separated equations related to Jacobi polynomials or even exceptional orthogonal polynomials. Previously most of the works were based on co-algebra or tensor product approaches, and our new approach can be greatly useful to the understanding of $D$-dimensional superintegrable systems. Another main result of this paper is the construction of universal quadratic algebras for these models with arbitrary functions.

We have presented two new families of superintegrable systems with partition of coordinates, generalizing the ones studied in \cite{21} and \cite{Faz}. By using the separation of variables realized by blocked spherical coordinates, both systems can be transformed to familiar forms but parametrized by operators which are central elements. These central elements can be regarded as constants, since they are mutually commute and commute with other differential operators in radial variables. This point is crucial, because it allows us to take advantage of some known results to obtain the quadratic algebras structures of the new superintegrable systems with partitioned coordinates. 

Let us remark that our Hamiltonians (\ref{sys1}) and (\ref{sys2a}) have classical analogs. As mentioned in the introduction, the method of subgroup separation of variables was used in \cite{Kal10} to obtain a family of classical superintegrable systems.  However, our method of block separation of variables is different from the one used in \cite{Kal10}. Our model Hamiltonians are minimally superintegrable in general, and the models in \cite{Kal10} seem to correspond to an extension of the special case of our (\ref{sys1}) in which each block contains one coordinate. 

As the double singular oscillator Hamiltonians (i.e the $N=2$ special cases of our models) are related via the Hurwitz transformation to systems involving monopoles \cite{Faz}, it would be interesting to find out whether or not the $N$-singular models presented in this paper are dual to any kind of generalized monopole systems. This question is under investigation and results will be published elsewhere. 

\section*{Acknowledgement}
IM was supported by by Australian Research Council Discovery Project DP160101376 and Future Fellowship FT180100099. YZZ was supported by Australian Research Council Discovery Project DP190101529 and National Natural Science Foundation of China (Grant No. 11775177).

\appendix
\section{Appendix}
In this Appendix, we present the following proposition which can be used to derive the quadratic algebra relations (\ref{quadratic-alg1}).
\begin{mypro}\label{pop}
Consider the following operators $\mathcal{H}, \mathcal{H}_1,\mathcal{H}_2$, $\mathcal{Z}$:
\begin{equation*}
\begin{aligned}
    &\mathcal{H}=\mathcal{H}_1+\mathcal{H}_2,\\
    &\mathcal{H}_1=-\frac{\partial^2}{\partial x^2}+\omega^2 x^2+\frac{g_1}{x^2},~~~\mathcal{H}_2=-\frac{\partial^2}{\partial y^2}+\omega^2 y^2+\frac{g_2}{y^2},\\
    &\mathcal{Z}=\bigg(x\frac{\partial}{\partial y}-y\frac{\partial}{\partial x}\bigg)^2-(x^2+y^2)\bigg(\frac{g_1}{x^2}+\frac{g_2}{y^2} \bigg)\\
    &~~=(x^2+y^2)\bigg(\frac{\partial^2}{\partial x^2}+\frac{\partial^2}{\partial y^2} \bigg)-\bigg(x\frac{\partial}{\partial x}+y\frac{\partial}{\partial y} \bigg)^2-(x^2+y^2)\bigg(\frac{g_1}{x^2}+\frac{g_2}{y^2} \bigg),
\end{aligned}
\end{equation*}
where $g_1,g_2$ are real parameters. Then it can be shown by direct computation that these operators satisfy the quadratic algebraic relations,
\begin{equation*}
\begin{aligned}
    &\comm{\mathcal{Z}}{\mathcal{H}_2}=\mathcal{Y},\\
    &\comm{\mathcal{Z}}{\mathcal{Y}}=8\mathcal{Z}\mathcal{H}-8\acomm{\mathcal{Z}}{\mathcal{H}_2}+8(g_1-g_2+1)\mathcal{H}-16\mathcal{H}_2,\\
    &\comm{\mathcal{H}_2}{\mathcal{Y}}=-8\mathcal{H}\mathcal{H}_2+8\mathcal{H}_2^2-16\omega^2\mathcal{Z}-8\omega^2(2g_1+2g_2-1).
\end{aligned}
\end{equation*}
\end{mypro}


\begin{thebibliography}{}
\bibitem{mil} Miller Jr, W., Post, S., \& Winternitz, P. (2013). Classical and Quantum Superintegrability with Applications. J. Phys. A: Math. Theor. 46, 423001. 
\bibitem{kal} Kalnins, E.G., Miller Jr, W., \&  Post, S. (2013). Contractions of 2D 2nd order quantum superintegrable systems and the Askey scheme for hypergeometric orthogonal polynomials. SIGMA 9, 057. 
\bibitem{mil2} Miller Jr, W. (2014) The theory of contractions of 2D 2nd order quantum superintegrable systems and its relation to the Askey scheme for hypergeometric orthogonal polynomials. J. Phys. Conf. Ser. 512, 012012.
\bibitem{1} Abouamal, I., \& Winternitz, P. (2018). Fifth-order superintegrable quantum systems separating in Cartesian coordinates: Doubly exotic potentials. J. Math. Phys. 59, 022104.
\bibitem{2} Escobar-Ruiz, A.M., Vieyra, J.L., \& Winternitz, P. (2017). Fourth order superintegrable systems separating in polar coordinates. I. Exotic potentials. J. Phys. A: Math. Theor. 50, 495206.
\bibitem{3} Marquette, I., Sajedi, M., \& Winternitz, P. (2017). Fourth order superintegrable systems separating in Cartesian coordinates I. Exotic quantum potentials. J. Phys. A: Math. Theor. 50, 315201.
\bibitem{4} Post, S., Tsujimoto, S., \& Vinet, L. (2012). Families of superintegrable Hamiltonians constructed from exceptional polynomials. J. Phys. A: Math. Theor. 45, 405202.
\bibitem{5}  Kibler, M Lamot, G.H., \& Winternitz, P. (1992). Classical trajectories for two ring-shaped potentials. Int. J. Quant. Chem. 43, 625.
\bibitem{6} Nikitin, A. G. (2013). Superintegrable systems with arbitrary spin. Ukrainian J. Phys. 58, 1046.
\bibitem{7} Winternitz, P., \& Yurduşen, İ. (2006). Integrable and superintegrable systems with spin. J. Math. Phys. 47, 103509.
\bibitem{8} D'Hoker, E., \& Vinet, L. (1984). Supersymmetry of the Pauli equation in the presence of a magnetic monopole. Phys. Lett. B 137, 72.
\bibitem{9} Jackiw, R. (1980). Dynamical symmetry of the magnetic monopole. Ann. Phys. 129, 183.
\bibitem{10} Labelle, S., Mayrand, M., \& Vinet, L. (1991). Symmetries and degeneracies of a charged oscillator in the field of a magnetic monopole. J. Math. Phys. 32, 1516.
\bibitem{11} Mardoyan, L.G. (2002). Five-dimensional $su(2)$-monopole: continuous spectrum. Phys. Atom. Nucl. 65, 1063.
\bibitem{12} Wu, T.T., \& Yang, C.N. (1976). Dirac monopole without strings: monopole harmonics. Nucl. Phys. B 107, 365.
\bibitem{kal2} Kalnins, E.G., Kress, J.M., \& Miller Jr, W. (2005). Second order superintegrable systems in conformally flat spaces. 1. 2D classical structure theory. J. Math. Phys. 46, 053509.
\bibitem{kal3} Kalnins, E.G., Kress J.M., \& Miller Jr, W. (2005). Second order superintegrable systems in conformally flat spaces. 2. The classical 2D Stackel transform. J. Math. Phys., 46, 053510.
\bibitem{13} Kalnins, E.G., Kress, J. M., \& Miller Jr, W. (2013). Extended Kepler-Coulomb quantum superintegrable systems in three dimensions. J. Phys. A: Math. Theor. 46, 085206.
\bibitem{14} Kalnins, E.G., Kress, J.M., \& Miller Jr, W. (2006). Second order superintegrable systems in conformally flat spaces. IV. The classical 3D Stackel transform and 3D classification theory. J. Math. Phys.
47, 043514.
\bibitem{15} Kalnins, E. G., Miller, W., \& Post, S. (2010). Models for the 3D singular isotropic oscillator quadratic algebra. Phys. Atom. Nucl. 73, 359.
\bibitem{16} Escobar-Ruiz, M.A., \& Miller Jr, W. (2017). Toward a classification of semi-degenerate 3D superintegrable systems. J. Phys. A: Math. Theor. 50, 095203.
\bibitem{17} Tanoudis, Y., \& Daskaloyannis, C. (2011). Algebraic calculation of the energy eigenvalues for the nondegenerate three-dimensional Kepler-Coulomb potential. SIGMA 7, 054.
\bibitem{cap} Capel, J.J., Kress, J.M., \& Post, S. (2015). Invariant classification and limits of maximally superintegrable systems in 3D. SIGMA 11, 038.
\bibitem{das} Daskaloyannis, C., \& Tanoudis, Y. (2010) Quadratic algebras for three-dimensional superintegrable systems, Phys. Atom. Nucl. 73, 214.
\bibitem{bal} Ballesteros, A., Enciso, A.A., Herranz F.J., \& Ragnisco, O. (2009). Superintegrability on N-dimensional curved spaces: Central potentials, centrifugal terms and monopoles. Ann. Phys. 324, 1219.
\bibitem{gab} Gaboriaud, J., Vinet, L., Vinet, S., \& Zhedanov, A. The generalized Racah algebra as a commutant, e-print arXiv:1808.09518.
\bibitem{sh10} Schulze-Halberga, A. (2010) Intertwining relations and Darboux transformations for Schrödinger equations in $(n+1)$ dimensions, J. Math. Phys. 51, 033521.
\bibitem{ktv01} Kuru, S., Tegmen, A., \& Vercin, A. (2001) Intertwined isospectral potentials in an arbitrary dimension, J. Math. Phys. 42, 3344.
\bibitem{lrbeh16} Latini, D., Ragnisco, O., Ballesteros, A., Enciso, A.A., Herranz, F.J. \& Riglioni, D. (2016) The classical Darboux III oscillator: factorization, Spectrum Generating Algebra and solution to the equations of motion, J. Phys. Conf. Ser. 670, 012031.
\bibitem{cnd09} Calzada, J.A., Negro, J., Del Olmo, M.A. (2009). Intertwining Symmetry Algebras of Quantum Superintegrable Systems. SIGMA 5,	039.
\bibitem{Kal10} Kalnins, E.G., Kress, J.M., \& Miller, W. (2010). Families of classical subgroup separable superintegrable systems, J. Phys.: Math. Theor. 43, 092001.
\bibitem{22} Rodrıguez, M. A., \& Winternitz, P. (2002). Quantum superintegrability and exact solvability in n dimensions. J. Math. Phys. 43, 1309.
\bibitem{LMZ} Liao, Y., Marquette, I., \& Zhang, Y.-Z. (2018). Quantum superintegrable system with a novel chain structure of quadratic algebras. J. Phys. A: Math. Theor. 51, 255201.
\bibitem{lat} Latini, D. (2019) Universal chain structure of quadratic algebras for superintegrable systems with coalgebra symmetry.J. Phys. A: Math. Theor. 52, 125202.
\bibitem{m84} Miller, W. (1984). Symmetry and separation of variables. Cambridge University Press. 
\bibitem{k86} Kalnins, E.G. (1986). Separation of variables for Riemannian spaces of constant curvature. New York: Wiley.
\bibitem{fmsuw65} Fris, J., Mandrosov, V., Smorodinsky, Ya. A., Uhlır, M., \& Winternitz, P. (1965). On higher symmetries in quantum mechanics. Phys. Lett 16, 354.
\bibitem{svw67} Makarov, A.A.,  Smorodinsky, Ya. A., Valiev, K., \& Winternitz, P. (1967). A systematic search for nonrelativistic systems with dynamical symmetries. Il Nuovo Cimento A 52, 10611084.
\bibitem{e90} Evans, N.W. (1990). Superintegrability in classical mechanics. Phys. Rev. A 41, 5666.
\bibitem{kkm18} Kalnins, E.G., Kress, J.M., \& Miller, W. (2018). Separation of variables and Superintegrability: The symmetry of solvable systems, IOPscience.
\bibitem{18} Petrosyan, M. (2008). Four-dimensional double-singular oscillator. Phys. Atom. Nucl. 71, 1094.
\bibitem{19} Marquette, I. (2010). Generalized MICZ-Kepler system, duality, polynomial, and deformed oscillator algebras. J. Math. Phys. 51, 102105.
\bibitem{20} Marquette, I. (2012). Generalized five-dimensional Kepler system, Yang-Coulomb monopole, and Hurwitz transformation. J. Math. Phys. 53, 022103.
\bibitem{Faz} Hoque, F.M., Marquette, I., \& Zhang, Y.-Z. (2015). A new family of N dimensional superintegrable double singular oscillators and quadratic algebra $Q(3)$ $\oplus so(n) \oplus so(N-n)$. J. Phys. A: Math. Theor. 48, 445207.
\bibitem{21} Chen, Z., Marquette, I., \& Zhang, Y.-Z. (2019). Extended Laplace-Runge-Lentz vectors, new family of superintegrable systems and quadratic algebras. Ann. Phys. 402, 78.
\bibitem{Cal} Calogero, F. (1969). Solution of a three-body problem in one dimension. J. Math. Phys. 10, 2191.
\bibitem{KYR} Kumari, N., Yadav, R. K., Khare, A., \& Mandal, B.P. (2017). A class of exactly solvable rationally extended Calogero–Wolfes type 3-body problems. Ann. Phys. 385, 57.
\bibitem{GKM} G\'omez-Ullate, D., Kamran, N., \& Milson, R. (2009). An extended class of orthogonal polynomials defined by a Sturm–Liouville problem. J. Math. Anal. Appl. 359, 352.
\end{thebibliography}
\end{document}